\documentclass[aps,prd,nofootinbib,twocolumn,superscriptaddress,preprintnumbers,balancelastpage,longbibliography,nobibnotes]{revtex4-2}

\usepackage{amsmath,amssymb} 
\usepackage{graphicx}
\usepackage{hyperref}
\usepackage{enumitem}
\usepackage{xcolor}
\usepackage{tikz-cd}
\usepackage{graphicx}
\usepackage{graphics}
\usepackage{blindtext}
\usepackage{hyperref}
\usepackage{subcaption}
\usepackage[compat=1.1.0]{tikz-feynman}
\usepackage{tikz} %This package allows to generate graphics on LaTeX
\usetikzlibrary{arrows} %Allows you to include different types of arrows in your tikz diagrams
\usetikzlibrary{decorations.markings} %Allows you to use decoration and make markings on TikzFigures

\tikzset{
	fermion/.style={postaction={decorate}, 
	   decoration={markings,mark=at position .575 with {\arrow{triangle 45}}}}, % 0.55 here means the position of the tip of the arrow in the line i.e. almost the center
	 scalar/.style={dashed,postaction={decorate},
           decoration={markings,mark=at position .575 with }}
}

\def\O{\mathcal{O}}
\def\L{\mathcal{L}}

\def\hc{\text{h.c.}}
\def\U32{$U(3) \times U(2)$}

\definecolor{aqua}{rgb}{0.4, 0.6, 0.7}

\newcommand{\AddrIFIC}{%
  Instituto de F\'{i}sica Corpuscular, CSIC-Universitat de Val\`{e}ncia, 46980 Paterna, Spain}

\begin{document}

%-------------------
% Title and authors
%-------------------
\title{Non-anomalous axions: lessons from the Majoron}

\author{Antonio Herrero-Brocal}
\email{antonio.herrero@ific.uv.es}
\affiliation{\AddrIFIC, Valencia, Spain}

%-------------------
% Abstract
%-------------------
\begin{abstract}
We show that when an anomalous and a non-anomalous global symmetries are spontaneously and simultaneously broken, the resulting Nambu--Goldstone boson is associated with the non-anomalous one. Applied to the Majoron, this implies that its underlying symmetry is $B-L$ rather than $L$, naturally resolving the domain wall problem. This result further demonstrates that effective couplings of the form $a \, F \, \tilde{F}$ do not uniquely indicate an anomalous origin, allowing the detection of non-anomalous Nambu--Goldstone bosons in searches for axion--gauge boson interactions.
\end{abstract}

\maketitle

\section{Introduction}\label{sec:intro}
The observation of neutrino oscillations~\cite{Kajita:2016cak,McDonald:2016ixn} implies that neutrinos are massive, in clear tension with the Standard Model (SM), where neutrinos are exactly massless. This masslessness is not merely due to the absence of right-handed neutrinos, but rather to the presence of an accidental global custodial symmetry that forbids Majorana mass terms from appearing via quantum corrections. The existence of nonzero neutrino masses therefore constitutes one of the main motivations for physics beyond the SM (BSM). If neutrinos are of Majorana nature, then this custodial symmetry must be broken. A particularly well--motivated possibility is that a global custodial symmetry is broken spontaneously. In this case, Goldstone’s theorem predicts the existence of a physical Nambu–Goldstone boson (NGB) in the spectrum: the Majoron ($J$)~\cite{Chikashige:1980qk,Chikashige:1980ui,Schechter:1981cv,Gelmini:1980re,Aulakh:1982yn}.

In the SM, there are two accidental $U(1)$ global symmetries, usually associated with Lepton ($L$) and Baryon ($B$) numbers. Neutrinos are charged under $L$ while neutral under $B$. Hence, in the standard lore, $J$ is typically associated with the spontaneous breaking of $L$. Accordingly, in Majoron models one typically imposes $L$ as a global symmetry and constructs an $L$--invariant Lagrangian. Furthermore, in most cases, $B$ remains unbroken, making the Lagrangian equally invariant under any linear combination $\alpha L + \beta B$. Hence, the spontaneous breaking of $L$ also breaks $\alpha L + \beta B$ for all $\alpha \neq 0$. In principle, one might think that the spontaneous breaking could be associated with any of these combinations. However, this freedom would be inconsistent since in most Majoron models $L$ (or, more generally, $\alpha L + \beta B$) is anomalous, whereas the particular combination $B-L$ is free of gauge anomalies. If a global and free of gauge anomalies symmetry is spontaneously broken, the Majoron emerges as an exact NGB associated with an exact global symmetry. In contrast, the NGB associated to the Spontaneous Symmetry Breaking (SSB) of an anomalous global symmetry becomes a pseudo--Nambu--Goldstone boson (pNGB) because of the anomaly, i.e., an axion--like particle (ALP), whose mass is generated by anomaly--induced effects. Depending on the anomaly structure, this may lead to additional phenomenological or cosmological consequences, including the appearance of topological defects such as domain walls~\cite{Zeldovich:1974uw}.

In this work, we solve the apparent inconsistency by showing explicitly that the Majoron (or, more generically, Majorana neutrinos nature) is univoquely associated with $B-L$ rather than with $L$. This claim can be extended to other NGBs: if the Vacuum Expectation Value (VEV) of some complex scalar spontaneously breaks simultaneously an anomalous symmetry and a symmetry free of gauge anomalies, the NGB is associated with the latter. We will show that this leads to some relevant phenomenological consequences such as the protection against the domain wall problem.

The rest of the manuscript is organized as follows. In Sec.~\ref{sec:B-L}, we show the $B-L$ nature of the Majoron. We then generalize this result to any NGB and discuss the case of the QCD axion. The phenomenological consequences associated with domain wall formation and axion-like couplings are explored in Sec.~\ref{sec:Pheno}. Finally, we summarize and conclude in Sec.~\ref{sec:sum}.

\section{The Majoron as the $B-L$ NGB}\label{sec:B-L}

In the following, we will restrict ourselves to the case where $B-L$ is free of gauge anomalies. As we will demostrate, in this case the Majoron is univoquely associated with $B-L$. To show this claim, let us first define our framework:
\begin{itemize}
\item We consider a theory invariant under $L$ and $B$, with the Lagrangian given by $\L_{\text{inv}}$. This theory includes a scalar, $\phi$, charged under $L$ but uncharged under $B$. This scalar acquires a VEV leading to the generation of Majorana neutrino masses.
\item As usual, $B$ and $L$ symmetries (and the combination $B+L$) are anomalous under $SU(2)_L$ and $U(1)_Y$, whereas $B-L$ is free of gauge anomalies.
\item We add to $\L_{\text{inv}}$ explicit symmetry--breaking terms: one that breaks $L$ and $B$ while conserving $B-L$ ($\O_{B-L}$), one that breaks $L$ and $B$ while conserving $B+L$ ($\O_{B+L}$), one that breaks only $B$ and conserves $L$ ($\O_{L}$), and another that breaks only $L$ while conserving $B$ ($\O_{B}$).
\item Since $L$, $B$, $B+L$, and $B-L$ are now all explicitly broken, Goldstone's theorem no longer applies. Consequently, it should be possible to identify which explicit breaking term generates the Majoron shift-symmetry breaking, thereby determining the symmetry underlying the Majoron.
\end{itemize}
Once defined our framework, our full Lagrangian is given by 
\begin{equation} \label{eq:Lfull}
\L = \L_{\text{inv}} + c_{B-L}  \O_{B-L} + c_{B+L}  \O_{B+L} +c_B   \O_B + c_L  \O_L \, ,
\end{equation}
with $c_i$  the Wilson coefficients.

Let us first focus on $\L_{\text{inv}}$, which is explicitly symmetry--invariant. As usual when dealing with NGBs, we employ the polar parametrization for $\phi$,
\begin{align}
\phi = e^{i q_\phi \pi(x)} \rho \, ,
\end{align}
which allows one to trade the explicit $\pi(x)$ dependence for derivative interactions $\partial_\mu \pi(x)$, making manifest the shift symmetry, as expected for a NGB. This can be achieved by exploiting the invariance of the Lagrangian under the $U(1)$ transformation
\begin{align} \label{eq:symtransf}
\psi \to e^{i q_\psi \theta} \psi \, ,
\end{align}
for each field $\psi$ carrying charge $q_\psi$. Thus, by promoting the transformation parameter $\theta$ to a space–time dependent function, $\theta \to \pi(x)$, $\pi(x)$ can be removed from all terms except those containing derivatives, which then generate its characteristic derivative couplings.

For an exact symmetry, this is the whole story: the NGB exhibits an exact shift symmetry, which ensures its masslessness. However, even if the symmetry is exact at the classical level, it can be broken by quantum effects. If the symmetry is anomalous with respect to a given gauge group, the transformation in Eq.~\eqref{eq:symtransf} generates the well--known $\theta$ term~\cite{Fujikawa:1979ay,Fujikawa:1980eg},
\begin{equation}
\mathcal{L}_{\text{anom}} \propto \theta\, F^a_{\mu\nu} \tilde{F}^{a\,\mu\nu} \, ,
\end{equation}
where $F^a_{\mu\nu}$ and $\tilde{F}^{a,\mu\nu}$ denote the gauge field strength and its dual, respectively, associated with the anomalous gauge group. Replacing $\theta$ by $\pi(x)$ then yields a coupling that is entirely analogous to the QCD axion coupling to gluons. Thus, as for the QCD axion, the anomaly breaks the NGB shift symmetry, thereby allowing for mass generation.

Applying this to our case, using $L$ to transform the Lagrangian would generate an anomalous coupling between the Majoron and the Electroweak (EW) gauge bosons. However, performing the field redefinition with the anomaly-free combination $B-L$ generates no such term. In this way, one may regard the $B-L$ transformation as a combination of an $L$ transformation followed by a compensating $B$ transformation that removes the anomaly. The fact that the anomaly can be eliminated by a field redefinition demonstrates that it is spurious and therefore does not correspond to a physical source of shift-symmetry breaking.This observation already suggests that the Majoron should be associated with $B-L$ rather than with $L$ itself. We now examine what happens once the explicit symmetry-breaking operators in Eq.~\eqref{eq:Lfull} are included. Performing the anomaly-free $B-L$ rotation, the Lagrangian transforms as
\begin{align} \label{eq:Jtrans}
\L \to & \tilde{\L}_{\text{inv}}\left( \partial_\mu J \right) + c_{B-L} \, \O_{B-L} + c_{B+L} \, e^{i Q_{B+L} \pi(x)} \, \O_{B+L} \nonumber \\
&+ c_B \, e^{i Q_{B} \pi(x)} \, \O_B + c_L \, e^{i Q_{L} \pi(x)} \, \O_L \, .
\end{align}
The operator $\O_{B-L}$ is $B-L$ invariant and, therefore, it does not transform. However, all the rest transform according to the total $B-L$ charge of the operator, $Q_i$. These terms are breaking the shift symmetry for the Majoron, thus becoming a massive pNGB. We note that if we were to choose any other $\alpha L + \beta B$ combination to rotate the Majoron, we would have found the same anomaly that appears only with $\mathcal{L}_{\text{inv}}$, which we now understand to be spurious. Therefore, to avoid this spurious shift symmetry breaking term, we must use $B-L$. Let us then emphasize this result. If we consider a theory with $c_{B-L}=c_{B+L}=c_B=0$ and we only break $B$ explictly, even if $L$ is conserved, this term breaks the shift symmetry which protects the masslessness of the Majoron. On the other side, if we consider $c_{B+L}=c_B=c_L=0$, then, even if $\O_{B-L}$ is breaking $L$ explicitly, the Majoron is an exact NGB. Therefore, this uniquely determines that the Majoron is not the Nambu–Goldstone boson of $L$, but of $B-L$.

The discussion here is of course more general than the particular case of the Majoron and $L$ or $B-L$. If the VEV of some complex scalar spontaneously breaks simultaneously an anomalous symmetry and a symmetry free of gauge anomalies, the NGB is associated with the non--anomalous one. This statement is particularly transparent when the symmetry is completely anomaly free, namely free of gauge, gravitational, and cubic anomalies. In that case, the global symmetry can be consistently promoted to a gauge symmetry. The corresponding NGB is then eaten by the gauge boson, becoming its longitudinal polarization. Since gauge invariance is exact, the associated shift symmetry must also be exact. Furthermore, the result can be also understood directly from Goldstone's theorem. The theorem states that every spontaneously broken generator of an exact global continuous symmetry gives rise to a NGB. In our setup, the only exact global continuous symmetry is the non-anomalous one, since the anomalous symmetry is explicitly broken by the anomaly. Therefore, Goldstone's theorem already implies that the NGB must be associated with the non-anomalous symmetry.

This general result also clarifies why explicit $B$-breaking generates a mass for the Majoron. Once $B$ is explicitly broken, no exact global symmetry remains in the theory: $B$ is violated at the Lagrangian level, while $L$ is broken by anomalies. Consequently, the assumptions of Goldstone’s theorem are no longer satisfied, and the Majoron acquires a mass. Moreover, in the absence of any exact symmetry, a dual description emerges in which all spontaneously broken symmetries can be treated on equal footing. We now illustrate this using the Majoron model with explicit $B$-breaking: 
\begin{equation}\label{eq:LB}
\mathcal{L}_B = \mathcal{L}_{\text{inv}} + c_L \, \mathcal{O}_L .
\end{equation}
We can then analyze the situation from two complementary perspectives:
\begin{itemize}
\item Non--anomalous picture. 

This is the framework used in the derivation of Eq.~\eqref{eq:Jtrans}. In this description, the operator $\mathcal{O}_L$ explicitly breaks the Majoron shift symmetry, showing that the Majoron is not an exact NGB. However, at the perturbative level, the theory still possesses an accidental symmetry identified with $L$, which appears to protect the NGB nature. Since this accidental symmetry is broken only non-perturbatively by anomalies, all effects that violate the shift symmetry must arise non-perturbatively, including mass generation. Importantly, however, these processes (including mass generation) necessarily arise because the shift symmetry has already been explicitly broken by $\mathcal{O}_L$.

\item Anomalous picture.

In this alternative description, the Majoron becomes derivatively coupled under an $L$ field redefinition. The operator $\mathcal{O}_L$ then appears invariant, while the anomaly becomes the only source of symmetry breaking. Since $B$ is explicitly broken, there is no field redefinition that can remove the shift-symmetry breaking (using $B$ just move the breaking from the anomaly to $\mathcal{O}_L$, recovering the non-anomalous picture previosly discussed). Moreover, since both $L$ and $B$ (and therefore also $B+L$) are broken ---explicitly or anomalously--- the electroweak vacuum angle $\theta_{\rm EW}$ becomes a physical parameter~\cite{FileviezPerez:2014xju,Anselm:1993uj,Anselm:1992yz}. In close analogy with the QCD axion, the anomaly induces an effective coupling between the Majoron and the electroweak gauge fields. Non-perturbative electroweak effects then generate a potential that dynamically relaxes $\theta_{\rm EW}$ to zero and simultaneously induce a mass for the Majoron.
\end{itemize}

Thus, once the non-anomalous symmetry is explicitly broken, both descriptions are physically equivalent. Indeed, this duality arises because both sources of symmetry breaking are required for the NGB to acquire a mass: the explicit breaking of $B-L$ and the anomalous breaking of $L$. This can be readily understood from the general result. In the limit $c_L \to 0$, $B-L$ becomes an exact, non-anomalous symmetry, and hence the NGB is an exact NGB. On the other hand, if $c_L \neq 0$ but the $L$ anomaly vanishes, then $L$ becomes a symmetry free of gauge anomalies. In this case, our general result shows that the NGB must be associated with $L$ being, again, massless. This can be made even more explicit by returning to the formalism introduced above. Consider a theory with two global $U(1)$ symmetries, $X$ and $Y$, such that $X$ is classically conserved but anomalous, whereas $Y$ is free of gauge anomalies but explicitly broken by an operator $\O_X$ that violates $Y$ while preserving $X$. Then, for the following Lagrangian 
\begin{equation}
\mathcal{L} = \mathcal{L}_{\text{inv}} + c_X \, \O_X \, ,
\end{equation}
it is clear that both sources of symmetry breaking (the anomaly and $\O_X$) are required to generate a mass for the NGB. If either of them is absent, one can always perform a transformation such that the NGB retains an exact shift symmetry.
This also allows us to clarify the nature of the Majoron. Throughout the derivation, we have worked within a framework in which both $L$ and $B$ are anomalous, whereas $B-L$ is anomaly-free. For this reason, we identified the Majoron with the $B-L$ symmetry. However, our general result shows that the Majoron should in fact be associated with the anomaly-free combination, whatever its precise form may be. In general, the anomaly-free symmetry could be any other linear combination.
%%%%%%%%%%%%%%%%%%%%%%%%
%%%%%%%%%%%%%%%%%%%%%%%%
\subsection{The QCD axion parallelism} \label{subsec:axion}
Finally, we can apply our result to the QCD axion case. The QCD axion is introduced to solve the strong CP problem~\cite{Weinberg:1977ma,Wilczek:1977pj,Peccei:1977hh}. In the standard picture, the QCD axion is the pNGB associated to the anomalous Peccei--Quinn (PQ) symmetry. However, once one takes into account the axial $U(1)$ symmetry, one realizes that the theory contains two $SU(3)_c$-anomalous symmetries: the PQ symmetry and the axial symmetry (the latter also being explicitly broken by the quark masses). Consequently, there exists a linear combination of both symmetries that is free of the $SU(3)_c$ anomaly. Applying the discussion above, we observe that the QCD axion admits two equivalent descriptions: the traditional one, in which it is associated with the anomalous PQ symmetry, and another in which the underlying symmetry is the combination of the axial and PQ symmetries that is free of the $SU(3)_c$ anomaly and is explicitly broken by the quark masses. This can be seen easily by using the QCD Lagrangian,
\begin{equation}
\mathcal{L}_{\text{QCD}} \supset
\sum_{f=1}^{N_f} \bar{q}_f \left( i \gamma^\mu D_\mu - m_f \right) q_f
+ \frac{\theta\, g_s^2}{32\pi^2}\, G^a_{\mu\nu} \tilde{G}^{a\,\mu\nu} \, ,
\end{equation}
and taking the axial transformation $q_f \to e^{i \frac{\theta}{2 N_f} \gamma_5} q_f$ to move the $\theta$ dependence into the quark masses,
\begin{equation}
\mathcal{L}_{\text{QCD}} \supset
 \sum_{f=1}^{N_f} \bar{q}_f \left( i \gamma^\mu D_\mu - m_f e^{i \frac{\theta}{N_f}} \right) q_f \, .
\end{equation}
Thus, if we include new fields to introduce the PQ symmetry and the axion solution, we no longer need to refer to the $SU(3)_c$ anomaly directly, but rather to the $SU(3)_c$ anomaly--free combination that is explicitly broken by the quark masses. In this picture, its mass arises from the interplay between the $SU(3)_c$ anomaly and the explicit breaking induced by $m_f$, just as for the Majoron discussed above.
Furthermore, it makes explicit, without the need for any calculation, that the axion mass vanishes in the limit $m_f \to 0$. In this limit, the anomaly-free symmetry becomes exact, and the axion, as its associated NGB, must therefore remain massless, even though the $SU(3)_c$ anomaly of the PQ symmetry is still present. One could argue that, in the limit $m_q \to 0$, the $\theta$ angle becomes unphysical, and that this is why the axion remains massless. Our framework, however, shows that the masslessness of the QCD axion is independent of the $\theta$ angle. To make this explicit, it is sufficient to introduce an operator $\mathcal{O}$ that explicitly breaks both the axial and PQ symmetries while preserving their anomaly-free combination. In this scenario, even in the limit $m_q \to 0$, the $\theta$ angle remains physical, yet the axion is still massless since it is associated with an exact anomaly-free symmetry.

Nevertheless, for the QCD axion the non--anomalous symmetry is explicitly broken and then the distinction is merely terminological. However, as we will see, since $B-L$ is an exact symmetry, this is no longer true for the Majoron and the general claim leads to important phenomenological consequences.

\section{Phenomenological consequences}\label{sec:Pheno}
The first and most immediate consequence is the exactly masslessness of the Majoron in the absence of explicit $B-L$ breaking. However, from both the model-building and experimental points of view, there are two particularly important consequences associated with the non-anomalous origin of the Majoron, on which we will focus.
%%%%%%%%%%%%%%%%%%%%%%%%
%%%%%%%%%%%%%%%%%%%%%%%%
\subsection{Domain walls}
There has been recent confusion regarding whether certain Majoron models, such as the singlet model, suffer from a domain wall problem or not (see, for example, these papers on the topic~\cite{Lazarides:2018aev,Brune:2022vzd,Brune:2025zwg,Berbig:2025nrt,Brune:2025ktu}). Domain walls are two--dimensional topological defects that arise when a discrete symmetry is spontaneously broken, leading to a set of physically distinct degenerate vacua. In many models, this discrete symmetry originates from the explicit breaking of an underlying continuous symmetry. Then, the same potential responsible for lifting the continuous symmetry generates a mass for the associated pNGB.

In the Majoron scenario, since $L$ is explicitly broken by the anomaly, the SSB of $L$ does not correspond to the breaking of a $U(1)$ but rather to that of a residual $\mathbb{Z}_n$, with $n$ depending on the anomaly coefficient. Then, if the Majoron were associated to $L$, it would be tied to a discrete symmetry, potentially leading to the formation of domain walls, as noted in~\cite{Lazarides:2018aev}. Instead, some authors have argued that the Majoron avoids this problem as long as there is no $L$ and $B$ explicit breaking in the combination $B+L$~\cite{Berbig:2025nrt,Heeck:2019guh}. The argument is that in such a case there is no physical $\theta_{\text{EW}}$ dependence and therefore $SU(2)$ instantons cannot generate a potential for the Majoron. However, this argument appears to be an accidental consequence of a symmetry unrelated to the Majoron nature.

As we have seen in this work, even in presence of a physical $\theta_{\text{EW}}$ (for example, via operators such as $QQQL$~\footnote{This fact is explicitly showed in the updated version of Ref.~\cite{Berbig:2025nrt}; see the acknowledgements of that version, where the authors kindly acknowledge discussions with us on this point.} which break $L$ and $B$ explicitly), the Majoron remains massless as long as $B-L$ is conserved.  This is a consequence of the Majoron independence of the anomaly. Hence, since $B-L$ is conserved by non--perturbative effects, Majoron models are safe from the formation of domain walls provided that there is not $B-L$ explicit breaking.

%%%%%%%%%%%%%%%%%%%%%%%%
%%%%%%%%%%%%%%%%%%%%%%%%
\subsection{Axion--like couplings}
As discussed above, the pNGB associated with an anomalous global symmetry exhibits a characteristic anomalous coupling to gauge bosons of the form
\begin{equation}
\L_{\text{anom}} \propto a F_{\mu \nu}^a \tilde{F}^{a \mu \nu} \, ,
\end{equation}
with $a$ the pNGB or ALP. The search for these couplings is a primary experimental focus (see for instance~\cite{Graham:2015ouw,Irastorza:2018dyq} for reviews). Typically, the observation of such an interaction is associated with a topological coupling that would identify an ALP and, thus, a Dark Matter (DM) candidate~\cite{Peccei:1977hh,Preskill:1982cy,Abbott:1982af} which potentially solves the Strong CP problem (if the ALP is the QCD axion)~\cite{Weinberg:1977ma,Wilczek:1977pj,Peccei:1977hh}. Furthermore, it would provide a test for GUTs, where gauge group unification implies fundamental relations between the various gauge boson couplings~\cite{Agrawal:2022lsp}. 

In this work we have shown that the Majoron is associated with the non-anomalous $B-L$ symmetry and therefore these anomalous couplings between the Majoron and electroweak gauge bosons do not arise. However, our general result has implications that go beyond this specific observation. Indeed, we have shown that the existence of these anomalous terms in the Lagrangian depends on the chosen symmetry description. For instance, in the case of the QCD axion, we have shown that one can work with a description where the QCD axion is associated with a linear combination of the axial and PQ symmetries that is free of the $SU(3)_c$ anomaly, and that this formulation is fully equivalent to the conventional one. In this basis, the axion does not exhibit an anomalous coupling to gluons. In the standard description, however, the axion couples to $G\tilde{G}$ through the $SU(3)_c$ anomaly, giving rise to an interaction that remains finite in the limit $m_a \to 0$, is independent of fermion masses, and is quantized. These properties are characteristic of topological terms. Nevertheless, both formulations are equivalent and must reproduce identical physical amplitudes, implying that the anomalous interaction is recovered dynamically. Consequently, although the anomalous description correctly captures the infrared structure of the theory, the coupling itself does not necessarily originate from a gauge anomaly. This shows that topological-looking interactions do not provide a unique diagnostic of anomalous symmetries.

This conclusion, which follows directly from our general result, further implies that topological-looking interactions are not a unique signature of anomalous symmetries, and that such interactions may also arise for non-anomalous axion-like particles, i.e. exact NGBs. Moreover, this is precisely what occurs in the type-I seesaw framework with SSB, where interactions such as $J Z \tilde{Z}$ can be generated without relying on any electroweak anomaly and independently of the Majoron mass~\cite{Heeck:2019guh}. This equivalence, which arises naturally in our framework by associating the NGB with a non-anomalous symmetry, was explicitly demonstrated in~\cite{Quevillon:2019zrd}. More generally, it implies that the Majoron (and, more broadly, NGBs of non-anomalous symmetries) can be probed experimentally through $a F\tilde{F}$ or $a G\tilde{G}$ searches, even in the absence of electromagnetic or $SU(3)_c$ anomalies.
%%%%%%%%%%%%%%%%%%%%%%%%
%%%%%%%%%%%%%%%%%%%%%%%%

Given the importance of this result, let us discuss it in detail for the case of the QCD axion. As is well known, the interaction of the QCD axion with gluons is independent of the quark masses. Therefore, one can take the $m_q \to 0$ limit while keeping the coupling fixed. However, as shown in Sec.~\ref{subsec:axion}, an exact symmetry emerges in the $m_q \to 0$ limit that is not broken by the $SU(3)_c$ anomaly. The axion is therefore an exact NGB associated with this $SU(3)_c$-anomaly-free symmetry, while nevertheless exhibiting a coupling of the form $a G\tilde{G}$.\\
To further illustrate how such interactions arise, we now consider a simple toy model in analogy with the QCD axion example, in which an anomaly-like $\pi F\tilde{F}$ coupling is generated at one loop for a non-anomalous NGB, even in the absence of an electromagnetic anomaly.
%%%%%%%%%%%%%%%%%%%%%%%%
%%%%%%%%%%%%%%%%%%%%%%%%
\subsubsection{Toy model}
Let us consider an extension of the SM in which we introduce four  fermions: $\psi_{L,R}$ transforming as $(1,1,2)$ under the SM gauge group and carrying $(B-L)\,(\psi_{L}) = 1$ and $(B-L)\,(\psi_R) = -1$, and $\chi_{L,R}$ transforming as $(1,1,1)$ under the SM gauge group and carrying $(B-L)\,(\chi_L) = -4$ and $(B-L)\,(\chi_R) = 4$. In addition, we introduce a scalar that transforms as a singlet under the SM gauge group while carrying $(B-L)\,(\phi)=2$.\\
The charges of $\chi_{L,R}$ are chosen such that $B-L$ remains free of gauge anomalies. To avoid any ambiguity, we additionally introduce an operator that violates $B$ and $L$ while preserving $B-L$, denoted by $\O$, such that $\theta_{EW}$ becomes physical. By imposing $B-L$ as a global symmetry, the Yukawa sector is then given by
\begin{align}
\L_Y &= \L_Y^{\textup{SM}} + \L_Y^{\textup{BSM}} + \O \, ,
\end{align}
with
\begin{align}
\L_Y^{\textup{BSM}} =  y \, \phi \, \bar{\psi}_L \, \psi_R + \hc\, .
\end{align}
The scalar potential satisfies $V(H, \phi) = V(H H^\dagger, \phi \phi^\dagger)$. Then, in the linear parametrization, the NGB is given by the CP-odd component of $\phi$,
\begin{align}
\phi = \frac{1}{\sqrt{2}} \left( v_\phi + \phi^0 + i \, \pi \right) \, ,
\end{align}
and is massless and associated to $B-L$.\\
On the other hand, in this framework, the $\pi \gamma \gamma$ amplitude arises at one loop through the diagram shown in Fig.~\ref{fig:agg}.
\begin{figure}[h!]
\centering
\begin{tikzpicture}[scale=0.7,
  >=stealth',
  pos=.8,
  photon/.style={decorate,decoration={snake,post length=1mm}}]
\draw[scalar] (-14,0)--(-12,0);
\node at (-14.5,0){$\pi$};
\draw[fermion, thick] (-12,0)--(-10,2);
\node at (-11.2,2){$f_i$};
\draw[fermion, thick] (-10,-2)--(-12,0);
\node at (-11.2,-2){$f_i$};
\draw[fermion,thick] (-10,2)--(-10,-2);
\node at (-9,0){$f_i$};
\draw[photon] (-8,2)--(-10,2);
\node at (-7.5,2){$\gamma$};
\draw[photon] (-8,-2)--(-10,-2);
\node at (-7.5,-2){$\gamma$};
\end{tikzpicture}
\caption{Feynman diagram leading to the 1--loop coupling of the NGB to a pair of photons. The index $i$ runs over all charged fermions of the theory.}
\label{fig:agg}
\end{figure}
Since $B-L$ is free of gauge anomalies, anomaly-based arguments would suggest that, although effective interactions of the form $\partial^n \pi \, F \, \tilde{F}$ are expected to appear~\cite{Nakayama:2014cza}, the $\pi \, F \, \tilde{F}$ coupling should vanish in the $m_\pi \to 0$ limit. This conclusion follows from adopting the polar parametrization for the fields according to their $B-L$ charges. Since $B-L$ is free of gauge anomalies, no anomalous term is generated. However, in the $m_\pi \to 0$ limit, the contribution of each charged fermion $f_i$ to the amplitude is given by~\cite{Quevillon:2019zrd}
\begin{equation}
\mathcal{M}_i =  -i \frac{ q_i^2 \, e^2}{4 \pi^2}\, \frac{Y_{ii}}{\left(\hat{M}_f\right)_{ii}} \, \epsilon^{\alpha \beta \rho \sigma} \, \epsilon^*_\alpha(p_1) \, \epsilon^*_\beta(p_2) \, p_{1,\rho} \, p_{2, \sigma} \, ,
\end{equation}
where $p_{1,2}$ denote the photon momenta, $q_i$ is the electric charge of $f_i$ in units of $e$, and $\left(\hat{M}_f\right)_{ii}$ is the mass of $f_i$. On the other hand, $Y_{ii}$ denotes the Yukawa coupling between the NGB and the charged fermion $f_i$. In our model, the tree-level interaction of the NGB with the SM leptons and quarks vanishes identically. Therefore, the only contribution to the amplitude comes from the exotic $\psi$ fermions. The coupling is then given by
\begin{equation}
\L_{\pi \psi^+ \psi^-} =  \frac{i}{\sqrt{2}} \, y\, \pi \, \bar{\psi}_L \psi_R = i \, \frac{M_\psi}{v_\phi} \, \pi \, \bar{\psi}_L \psi_R \, , 
\end{equation}
and the amplitude for the $\pi \, F \, \tilde{F}$ becomes
\begin{equation} \label{eq:pigg}
\mathcal{M}_i =  -i \frac{ q_{\psi}^2 \, e^2}{4 \pi^2 \, v_\phi} \, \epsilon^{\alpha \beta \rho \sigma} \, \epsilon^*_\alpha(p_1) \, \epsilon^*_\beta(p_2) \, p_{1,\rho} \, p_{2, \sigma} \neq 0 \, .
\end{equation}
Hence, an exact massless NGB acquires a one-loop coupling to photons that is formally indistinguishable from the anomalous interaction $g_{\pi\gamma\gamma}\, \pi \, F_{\mu\nu} \, \tilde{F}^{\mu\nu}$, despite the absence of any underlying electromagnetic anomaly.\\
We can easily understand the result. The exactness (or the  anomaly freedom) and the mass of the NGB depends on all the charged fermions of the theory. In particular, it requires the presence of $\chi_{L,R}$ which are necessary to cancel the gauge anomalies. However, as we have explicitly shown, the $\pi \gamma \gamma$ coupling only depends on $\psi_{L,R}$. This fact demonstrates that the coupling must be independent of the anomaly.\\
In other words, the fact that $\chi_{L,R}$ do not interact with the NGB (and are therefore massless) makes them invisible to the NGB interactions in the linear parametrization. In this sense, in the linear parametrization the NGB only \emph{sees} the symmetry defined by $\psi_{L,R}$, with respect to which the rest of the fields are neutral. This symmetry is, of course, anomalous, since it does not include the contribution of $\chi_{L,R}$, necessary to cancel out the anomalies of the full symmetry. This fact explains why this triangle diagram, traditionally associated with anomalies, gives a non-zero result. The matching of the coupling with the anomaly given by this additional anomalous $U(1)$ is not a coincidence. We can see this from two alternative perspectives. First, let us discuss the role played by the masslessness of $\chi_{L,R}$. If $\chi_{L,R}$ were massive, then the mass term would break $B-L$. If the mass were explicit, $B-L$ would be explicitly broken, and, as discussed above, a duality between the anomalous and the non-anomalous description would emerge, explaining the result for the coupling in this case. Since this interaction is independent of $M_\chi$ in the explicit-breaking case, it must reduce to the same value in the $M_\chi \to 0$ limit, recovering exactly the massless scenario discussed above.~\footnote{This is completely equivalent to the QCD axion case discussed above.} If, instead, the mass were generated via SSB, $\chi_{L,R}$ would acquire a coupling to the NGB and would cancel the contribution of $\psi_{L,R}$, since now the NGB would \emph{see} the symmetry of the full sector $\chi_{L,R}$ and  $\psi_{L,R}$, which is free of gauge anomalies. This can be made explicit with the symmetry-preserving operator
\begin{align}
\Delta \L = \frac{\kappa}{\Lambda^3} \, (\phi^\dagger)^4 \, \bar{\chi}_L \chi_R + \hc \, ,
\end{align}
and noting that $Y_\chi / M_\chi = -4/v_\phi$, which exactly cancels the contribution of $\psi_{L,R}$.\\
As an alternative point of view to understand the matching, we can imagine the model after integrating out the $\psi_{L,R}$ fermions. The resulting EFT exhibits an anomalous $B-L$ symmetry (since in this EFT we only have the fermions from the SM and $\chi_{L,R}$) but with the presence of the $\pi \gamma \gamma$ coupling, as it is independent of the $\psi_{L,R}$ mass. Therefore, the fact that our original theory had a $B-L$ symmetry free of gauge anomalies must be translated into the EFT in the fact that, under a $B-L$ transformation, the anomaly induced by $\chi_{L,R}$ must be compensated by the transformation of the allowed dimension-5 operator $\pi F \tilde{F}$ (which is equivalent to saying that the anomaly of $\chi_{L,R}$ compensates that of $\psi_{L,R}$ in the complete theory). This simply reflects the fact that the anomalous term was indeed reabsorbable and therefore unphysical.\\
However, it is important to note that, even though the matching works for predicting the correct interaction, our general result shows that it does not come from the anomaly, as such. In fact, we know that, in the complete theory, this term is unphysical, since the symmetry behind the NGB is the gauge-anomaly-free $B-L$. This can be seen more explicitly in the case of the QCD axion discussed above. In the limit $m_f \to 0$ but with $\theta$ physical (for instance, by introducing an operator that violates both the $PQ$ and axial symmetries while preserving their combination), the $a G \tilde{G}$ coupling is generated, and the correct result can be recovered from the anomaly interpretation, exactly as we did above for our toy model. However, it is transparent in this case that the anomaly cannot be the real mechanism behind the interaction: if it were, one could not explain why the coupling is generated but not the mass. In our toy model we chose the simplest example, where the only relevant interaction is with the gauge bosons associated with a $U(1)$ symmetry, which admits no non-trivial instantons. However, if one were to build an equivalent toy model with $SU(2)$ doublets instead of singlets, it becomes clear, as in the QCD axion example, that the anomaly interpretation alone could not simultaneously explain both the presence of the coupling and the absence of a mass from $SU(2)$ instantons. This reinforces our result: in this alternative toy model, adopting the picture of the anomalous symmetry would give the correct result for the coupling; but if one instead believed the anomaly to be responsible for the interaction, one would also be forced to conclude that the NGB is massive and that the model suffers from the domain wall problem. Our result shows that neither of these is true. In such a toy model, the NGB is massless and associated with an exact symmetry free of gauge anomalies. Therefore, there is no domain wall formation, yet the axion-like coupling to gauge bosons remains finite.\\

Finally, let us emphasize that, although formally identical to the anomalous interaction, this effective coupling does not violate the shift symmetry of the NGB. Indeed, the general expression for the amplitude in Eq.~\eqref{eq:pigg} can be rewritten as
\begin{equation}
\mathcal{M} = -i \frac{q_i^2 \, e^2}{4 \pi^2}\, \frac{Y_{ii}}{\left(\hat{M}_f\right)_{ii}} \, \epsilon^{\alpha \beta \rho \sigma} \, \epsilon^*_\alpha(p_1) \, \epsilon^*_\beta(p_2) \, p_{1,\rho} \, p_{\pi, \sigma} \, ,
\end{equation}
where we have made use of the antisymmetry of the Levi-Civita tensor and momentum conservation  $p_\pi= p_1 + p_2$,  with $p_\pi$ the NGB momentum. This form makes the consistency with the NGB shift symmetry manifest, as the amplitude satisfies the Adler zero in the soft limit $p_\pi\to 0$~\cite{Adler:1965ga}.\\

\section{Summary}\label{sec:sum}
In Majoron models, the imposition of $L$ or $B-L$ often appears somewhat arbitrary, typically associating $J$ with the SSB of $L$. In this work we have shown that, when there is simultaneous breaking of an anomalous symmetry and another that is free of gauge anomalies, the NGB is associated with the non-anomalous one. For the Majoron, this means that its underlying symmetry (and therefore the one related to the Majorana nature of neutrinos) is $B-L$ rather than $L$. This resolves the domain wall problem in Majoron models even in the presence of a physical $\theta_{\text{EW}}$, provided there is no explicit breaking of $B-L$.

On the other hand, the search for ALPs is one of the most active experimental directions in particle physics. In particular, the observation of an ALP, understood as a pNGB associated with an anomalous global symmetry, would have important phenomenological implications, such as providing a candidate to address the strong CP problem, serving as a DM candidate, or even offering a means to test the viability of certain GUT scenarios through their predicted axion couplings with gauge bosons. Thus, determining when a pNGB is associated with an anomalous symmetry or not is of vital importance. In this work, using our main result, we have shown explicitly that the generation of physical processes mediated by the effective operator $a \, F \, \tilde{F}$ between a NGB and two gauge bosons is not uniquely associated with a NGB linked to an anomalous symmetry, thereby allowing the detection of non--anomalous NGBs in this type of experiment, potentially even serving as evidence for the existence of new charged fermions.

\begin{center}
\vspace*{0.5cm}
\textbf{Acknowledgements}
\end{center}

I am especially grateful to Avelino Vicente for helpful discussions and comments on the manuscript. I am also grateful to Maximilian Berbig and Mario Reig for discussions on related topics. I thank the anonymous Referee for helpful comments that improved the presentation and the results of this work, in particular for pointing out an issue with the original toy model. This work is supported by the Spanish grants
PID2023-147306NB-I00 and CEX2023-001292-S (MCIU/AEI/10.13039/501100011033), as
well as CIPROM/2021/054 and CIACIF/2021/100 (Generalitat Valenciana).

%-------------------
% Bibliography
%-------------------
\bibliographystyle{apsrev4-2}
\bibliography{refs.bib}

\end{document}